\documentclass[10pt,letterpaper]{article}
\usepackage[top=0.85in,left=1.in,footskip=0.75in,marginparwidth=2in]{geometry}

\usepackage[utf8]{inputenc}

\usepackage{cite}

\usepackage{nameref,hyperref}

\usepackage[right]{lineno}

\usepackage{microtype}

\raggedright
\usepackage{changepage}

\usepackage[aboveskip=1pt,labelfont=bf,labelsep=period,singlelinecheck=off]{caption}

\makeatletter
\renewcommand{\@biblabel}[1]{\quad#1.}
\makeatother

\usepackage{lastpage,fancyhdr,graphicx}
\usepackage{epstopdf}
\fancyheadoffset[L]{2.25in}
\fancyfootoffset[L]{2.25in}

\usepackage{color}

\definecolor{Gray}{gray}{.25}

\usepackage{graphicx}

\usepackage{sidecap}

\usepackage{tabularx}
\usepackage{mathrsfs}
\usepackage{lineno}
\usepackage{epsfig}
\usepackage{amsmath}
\usepackage{amssymb}
\usepackage{float}
\usepackage{enumerate}
\usepackage{lineno}
\usepackage{booktabs}
\usepackage{arydshln}

\usepackage{wrapfig}
\usepackage[pscoord]{eso-pic}
\usepackage[fulladjust]{marginnote}
\reversemarginpar

\begin{document}
\vspace*{0.35in}

\begin{flushleft}
{\Large
\textbf\newline{Control of Electron Energy Distribution Functions by Current Waveform Tailoring in Inductively Coupled Radio Frequency Plasmas}
}
\newline
\\
Zhaoyu Chen \textsuperscript{1},
Zili Chen \textsuperscript{1,*},
Yu Wang \textsuperscript{2},
Jonas Giesekus \textsuperscript{3},
Wei Jiang \textsuperscript{2},
Yonghua Ding \textsuperscript{1},
Donghui Xia \textsuperscript{1,*},
Ya Zhang \textsuperscript{4,*}
Julian Schulze \textsuperscript{3},
\\
\bigskip
\bf{1} State Key Laboratory of Advanced Electromagnetic Technology, International Joint Research Laboratory of Magnetic Confinement Fusion and Plasma Physics, School of Electrical and Electronic Engineering, Huazhong University of Science and Technology, Wuhan, 430074, China
\\
\bf{2} School of Physics, Huazhong University of Science and Technology, Wuhan 430074,China
\\
\bf{3} Chair of Applied Electrodynamics and Plasma Technology, Faculty of Electrical Engineering and Information Sciences, Ruhr University Bochum, 44801 Bochum, Germany
\\
\bf{4} School of Physics and Mechanics, Wuhan University of Technology, Wuhan 430070, China
\\
\bigskip
* zlchen@hust.edu.cn, xiadh@hust.edu.cn, yazhang@whut.edu.cn

\end{flushleft}

\section*{Abstract}

Based on two-dimensional particle-in-cell simulations a novel approach towards Electron Energy Probability Function (EEPF) and plasma chemistry control by Current Waveform Tailoring (CWT) in the coil of inductively coupled discharges is proposed. Varying the shape of this current waveform provides electrical control of the dynamics of the electric field in the plasma. Using sawtooth instead of sinusoidal waveforms allows breaking and controlling the temporal symmetry of the electric field dynamics. In this way CWT allows controlling the EEPF, the ionization-to-excitation rate ratio, and the plasma chemistry. 

\section{Introduction}

Technological high frequency low temperature plasmas represent complex multi-species non-linear physical systems of extreme societal relevance. They constitute a cornerstone of modern fabrication technologies, driving innovations across diverse fields ranging from semiconductor processing to environmental applications and biomedicine \cite{lieberman1994principles}. Nevertheless, the fundamental physical mechanisms of their operation remain poorly understood in many cases and, thus, knowledge based efficient plasma process control is limited. Such insights require a detailed understanding of the mechanisms of electron power absorption and the development of control concepts thereof to tailor the generation of distinct process relevant particle species generated by electron energy dependent dissociation and ionization of the background gas via EEPF control.

Recent advances have significantly deepened our understanding of the complex kinetic phenomena inherent in these discharges. Detailed investigations into sheath dynamics have revealed the critical role of secondary electron emission \cite{Sydorenko2009,Daksha_2019} and kinetic theory of emitting boundaries \cite{Sheehan2013}. Furthermore, the demonstration of similarity laws for discharge scaling \cite{Fu2025} and the enhancement of harmonic heating via the magnetized plasma series resonance \cite{Sun2024} highlight the rich landscape of physical mechanisms governing electron power absorption. Fundamental studies have also elucidated the complex structure of the EEPF and the abrupt transitions between heating modes \cite{Godyak1990,Schulze_2011x,Liu_2016}, underscoring the necessity for precise kinetic control.

Manipulation of the EEPF to selectively drive specific plasma chemical reaction pathways is a central goal, especially in capacitively coupled plasmas (CCP) and inductively coupled plasmas (ICP). In CCPs, the introduction of Voltage Waveform Tailoring (VWT) fundamentally altered the landscape of discharge control by providing electrical control of electron and ion energy distribution functions \cite{Buzzi_2009,Schulze_2011}. By exploiting the Electrical Asymmetry Effect \cite{Czarnetzki_2011,Lafleur_2016}, researchers achieved independent control over ion flux and energy, effectively breaking the constraints associated with conventional sinusoidal driving voltage waveforms. Other studies demonstrated control of the generation of distinct radicals by VWT in CCPs \cite{Wang_2024}. Additionally, kinetic phenomena such as collisionless bounce resonance resonance heating \cite{liu2011} and the non-linear enhancement of collisionless heating via multi frequency excitation \cite{Turner2006} have been identified as key drivers for electron energization in these systems.

However, despite these advancements in CCPs, no such electrical control concepts exist in ICPs, which constitutes a different type of discharge operated based on different power absorption mechanisms. Such plasmas are frequently used for plasma etching, where selective generation of radicals and ions is particularly important. Thus, developing such control concepts for ICPs would solve this critical problem by introducing a new technique of plasma control. This would open a new research avenue within low temperature plasma science, where the effects of different discharge conditions, gasses, reactor designs, etc. as well as the benefits of such control schemes for a variety of plasma processes could be explored.

Conventional ICPs are driven by sinusoidal coil currents \cite{turner1999hysteresis} and operate predominantly in an inductive H-mode \cite{hopwood1992review}, where electrons absorb power from the azimuthal induced electric field twice per RF period at equal strength. In low pressure regimes, electron heating in these discharges is governed by non local kinetic mechanisms \cite{cunge1999characterization} such as the anomalous skin effect \cite{Tyshetskiy2003}, which distinguishes them from collisional systems \cite{Turner1993}. Capacitive coupling  of the coil (E mode) can also contribute to electron power absorption during the phase of sheath expansion within each RF period.

We present a novel electrical EEPF control mechanism in ICPs based on tailoring the coil current waveform. It allows controlling the electron kinetics in the plasma and can be realized in applications by upgrading existing plasma sources by modifying only the external power supply based on existing multi-frequency impedance matchings \cite{Schmidt_2015,Schmidt_2018,Wang_2019} without changing the reactor itself. Utilizing 2D axisymmetric and experimentally validated Direct-Implicit Particle in Cell Monte Carlo Collision (PIC/MCC) simulations \cite{chenEHmode}, we demonstrate that and explain why non-sinusoidal sawtooth coil current waveforms with different and controllable rise-/fall-times provide electrical EEPF control.

\section{Computational model}

\begin{figure}[htbp]
	\begin{center}
		\includegraphics[width=\linewidth]{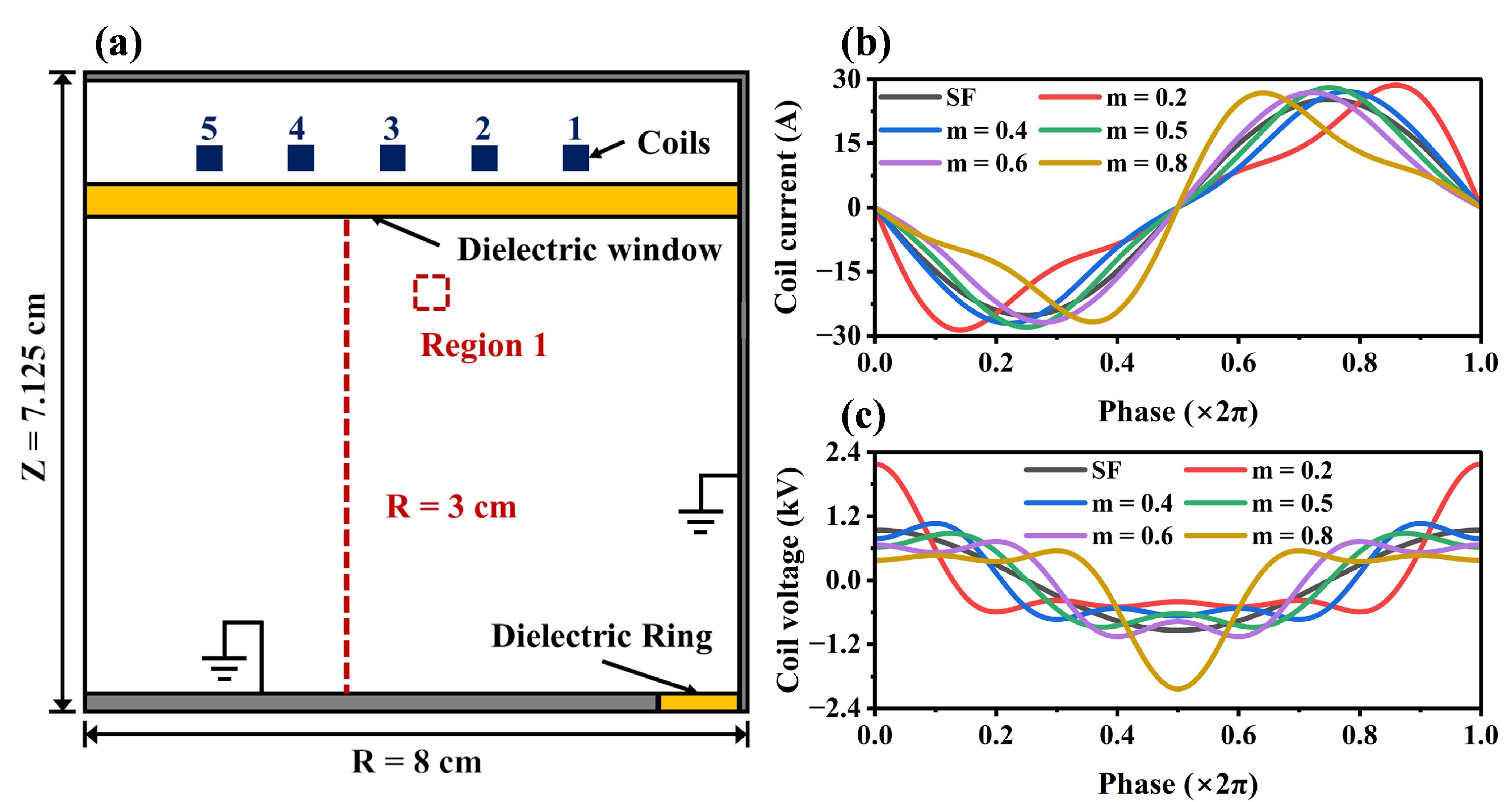}
	\end{center}
\caption{(a) 2D axisymmetric simulation domain showing the reactor geometry. Region 1 represents a $1 \times 1$ cm$^2$ square centered at $(r, z) = (4, 5)$ cm used for data analysis.  (b) Coil current and (c) corresponding coil voltage waveforms for the reference single frequency case (SF) and tailored coil current waveforms characterized by different m-values (equ. 1).}
\label{model}
\end{figure}

The simulation domain, shown in Fig. \ref{model}, represents a cylindrical reactor with a radius of 8 cm and a height of 5.8 cm. A grounded electrode is placed at the bottom and the reactor sidewall is grounded, too. The inductive source consists of a five-turn planar helical coil with a square cross-section, modeled as concentric rings. The simulations are performed in Argon at 20 mTorr and 300 K neutral gas temperature. The coil is driven by a tailored current waveform, $I(t)$, synthesized by the superposition of harmonics of a fundamental frequency, $\omega_0 = 2\pi \times 13.56$ MHz, up to the 4th order ($n=4$):
\begin{equation}
I(t)=I_0 \sum_{k=1}^n \frac{2(-1)^k}{k^2 m(1-m) \pi^2} \sin (k(1-m) \pi) \sin (k \omega_0 t),
\label{eq_waveform}
\end{equation}
where $m$ is an adjustable external parameter, that sets the amplitude of each harmonic. We compare the electron kinetics in the presence of 3 different current waveform shapes, i.e., the single frequency reference case (SF, $I(t) = - I_0 \sin (\omega_0 t)$) and 2 multi-frequency sawtooth waveforms (Sawtooth-Up: $m=0.2$, Sawtooth-Down: $m=0.8$).  In all these cases, $I_0$ is iteratively adjusted to maintain a fixed inductive power dissipation in the plasma of $P_{\mathrm{Ind}} = 80$ W within a tolerance of $\pm 2\%$. The resulting capacitive power absorption in the plasma, $P_{\mathrm{Cap}}$, is also evaluated and summarized in Table \ref{power}. For applications, the capacitive coupling of the coil can be minimized by placing a Faraday shield underneath the coil.

\begin{table}[htbp]
\centering
\renewcommand{\arraystretch}{2} 
\setlength{\tabcolsep}{5pt}      
\caption{Coil current $I_0$ required to maintain $P_{\mathrm{Ind}} \approx 80$ W and capacitive power dissipation, $P_{\mathrm{Cap}}$.}
\begin{tabular}{lcccccc}
\hline \hline
\textbf{Case (m)} & $\mathbf{SF}$ & $\mathbf{0.2}$ & $\mathbf{0.4}$ &
$\mathbf{0.5}$ & $\mathbf{0.6}$ & $\mathbf{0.8}$                          \\ \hline

$I_{\mathrm{0}}$ (A) & 25.11 & 30.67 & 30.76 & 30.99 & 30.49 & 28.67      \\ \hline

$P_{\mathrm{Ind}}$ (W) & 79.45 & 80.86 & 81.38 & 79.64 & 80.42 & 81.16    \\ \hline

$P_{\mathrm{Cap}}$ (W) & 13.89  & 32.24 & 12.88  & 14.01  & 12.74  & 18.12     \\ \hline

\hline
\end{tabular}
\label{power}
\end{table}

The model incorporates an energy-dependent description of electron induced secondary electron emission (SEE) from SiO$_2$ surfaces \cite{horvath2017role} and an ion-induced SEE coefficient of $\gamma_i=0.15$ \cite{lafleur2013secondary}.

To accurately resolve the induced electric field generated by the tailored coil current waveforms, we decompose this field into its harmonic constituents. The Helmholtz equation is solved separately for each harmonic order, $k$, to obtain the complex field amplitude, $\hat{E}_{\theta, k}$ \cite{takekida2006particle}:
\begin{equation}\begin{split}\left(\frac{\partial^2}{\partial r^2}+\frac{1}{r} \frac{\partial}{\partial r}+ \varepsilon_r \varepsilon_0 \mu_0 (k\omega_0)^2-\frac{1}{r^2}+\frac{\partial^2}{\partial z^2}\right) \hat{E}_{\theta, k}
\\
= i (k\omega_0) \mu_0 (\hat{J}_{\theta, k} + \hat{J}_{\mathrm{coil}, k})
\end{split}
\label{eq01}
\end{equation}
where $\hat{J}_{\theta, k}$ and $\hat{J}_{\mathrm{coil}, k}$ are the Fourier components of the plasma and coil current densities at the $k$-th harmonic, respectively. The total time-dependent electric field is then reconstructed via the superposition principle: 
$E_\theta(t) = \sum_{k=1}^n \hat{E}_{\theta, k} e^{j k \omega_0 t}$.

To capture capacitive coupling of the coil, the potential distribution along the coil is determined self-consistently. Invoking Faraday’s law in cylindrical symmetry, the instantaneous electromotive force (EMF) for the $i$-th turn, $\mathcal{E}_i(t)$, is computed by integrating the azimuthal electric field along the loop, while the potential relative to the grounded first turn is obtained by summation \cite{chenEHmode}:
\begin{equation}
\left\{
\begin{aligned}
& \mathcal{E}_i(t) = \oint \boldsymbol{E}_\theta(t) \cdot d \boldsymbol{l} = \int_{0}^{2\pi} r E_\theta(r, z, t) d\theta \\
& \mathcal{E}_{turn,N}(t) = \sum_{i=1}^N \mathcal{E}_i(t)
\end{aligned}
\right.
\label{eq_emf_potential}
\end{equation}

With the time resolved potential of each coil turn determined, the electrostatic field is calculated using an implicit formulation of Poisson's equation, simultaneously considering both metallic and dielectric materials \cite{wang2010implicitjiang}. The spatio-temporally resolved total electric field, i.e., the sum of the axial electrostatic and the azimuthal induced fields, is then used to accelerate and track charged particles in the simulation. Based on this kinetic approach the above equations are solved iteratively until convergence is reached.

\section{Results} 

While the inductively dissipated power is maintained at 80 W, while tailoring the coil current waveform, and is almost fully dissipated to electrons, the capacitive power dissipation in the plasma is lower and shows a maximum for the sawtooth-up current waveform (m = 0.2), while it is otherwise approximately constant (see table \ref{power}). A large fraction of $P_{\mathrm{Cap}}$ is dissipated to positive ions within the sheath adjacent to the coil. Its maximum for m = 0.2 is caused by the specific peaks coil voltage waveform required to realize the sawtooth-up coil current waveform, which leads to a long time of sheath expansion at the coil. Tailoring the coil current waveform fundamentally alters and provides electrical control of the temporal dynamics of the electric field and electrons in the plasma. 

\begin{figure}[htbp]
	\begin{center}
		\includegraphics[width=\linewidth]{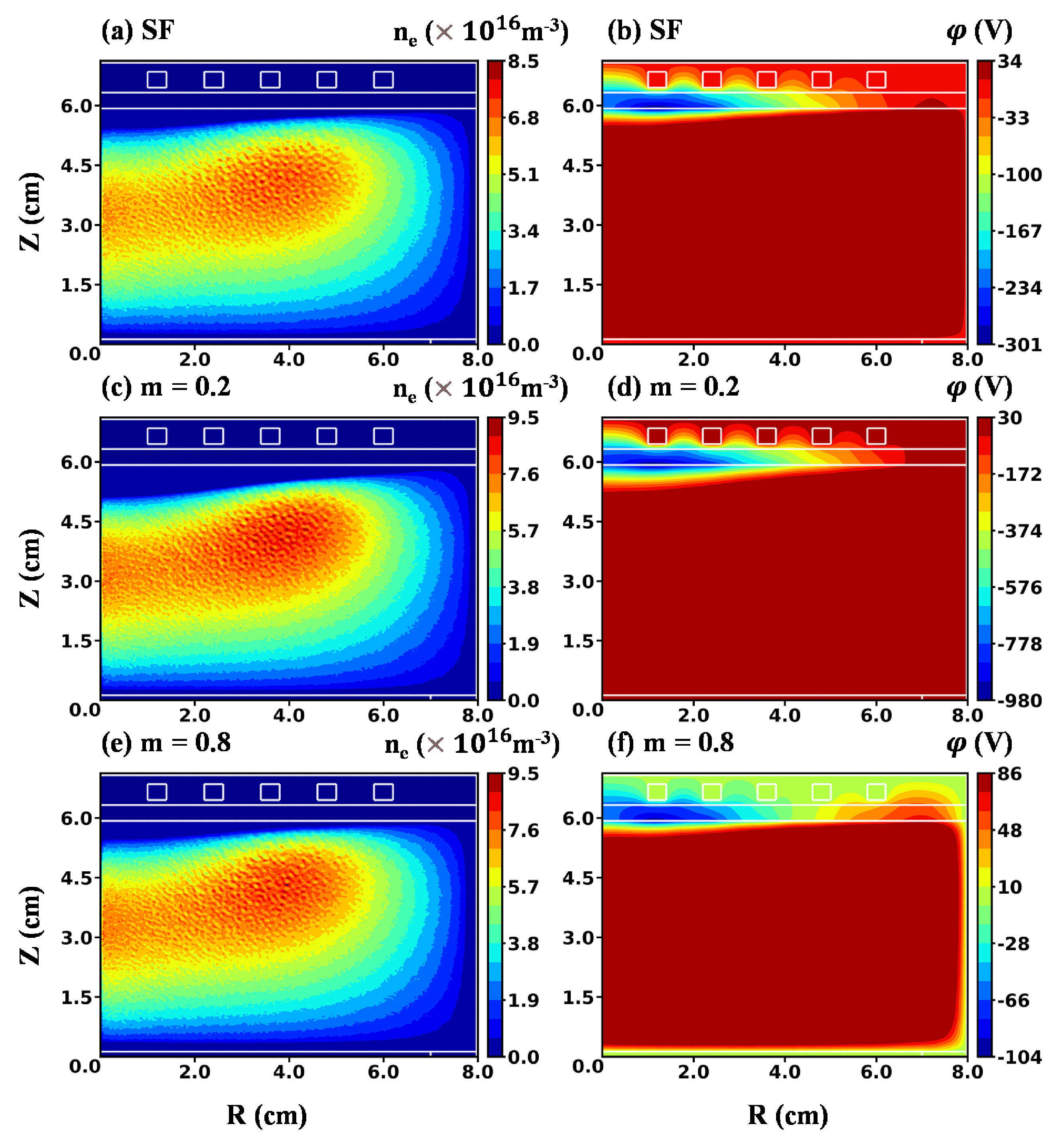}
	\end{center}
\caption{Spatial distributions of the (a, c, e) electron density, $n_e$, and (b, d, f) electric potential, $\varphi$, averaged over one fundamental RF period. }
\label{neandphi}
\end{figure}

\begin{figure*}[htbp]
\centering
\includegraphics[width=1\textwidth]{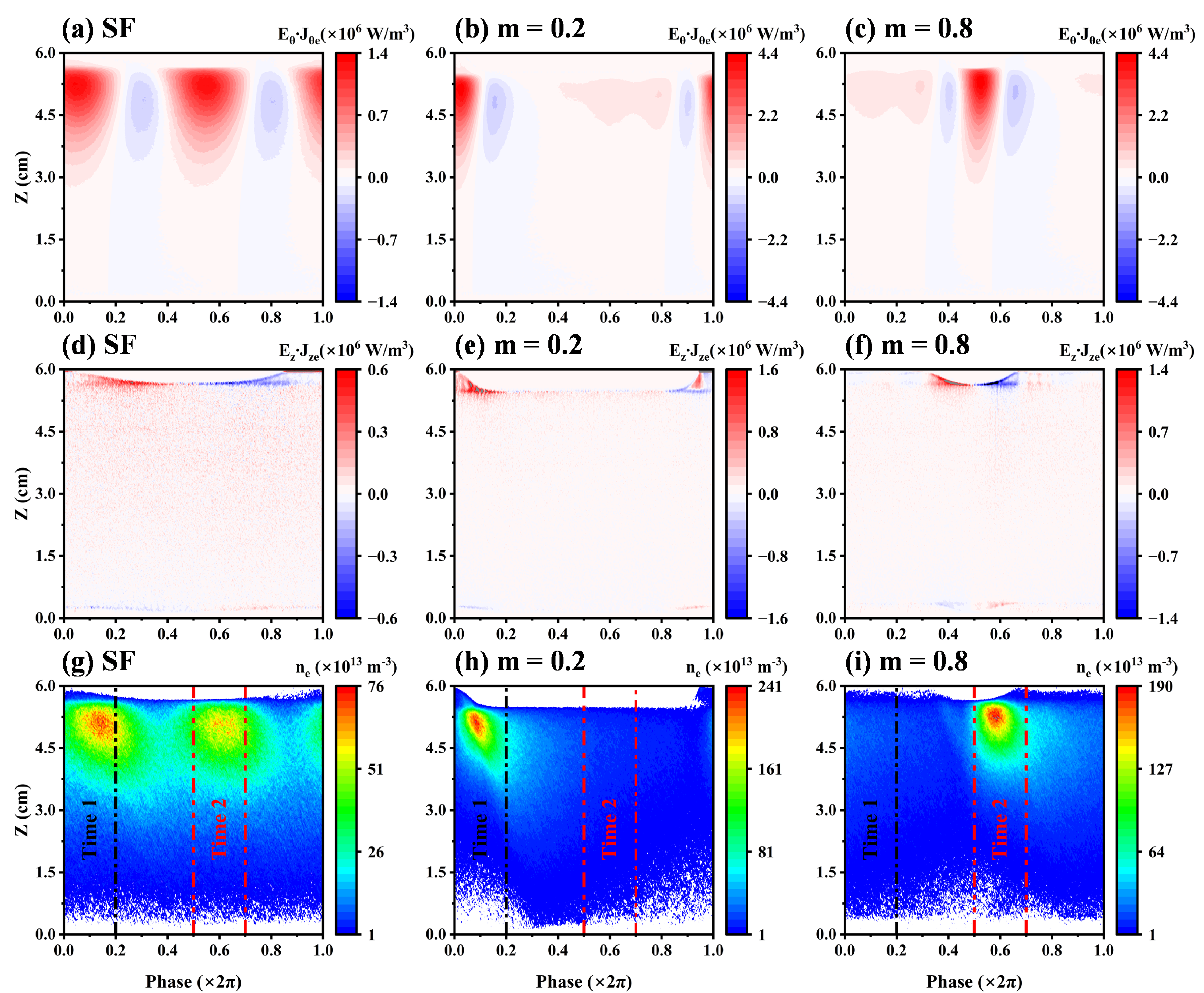}
\caption{Spatiotemporal evolution of electron dynamics at $R = 3$ cm over one fundamental RF period. The rows show: (a)–(c) azimuthal inductive heating rate $E_{\theta} \cdot J_{\theta e}$; (d)–(f) axial heating rate $E_{z} \cdot J_{z e}$; and (g)–(i) number density of high-energy electrons with energy $\ge 16~\mathrm{eV}$. The sampling windows for the EEPF are denoted by Time 1 and Time 2.}
\label{heating}
\end{figure*}

\begin{figure}[htbp]
	\begin{center}
		\includegraphics[width=\linewidth]{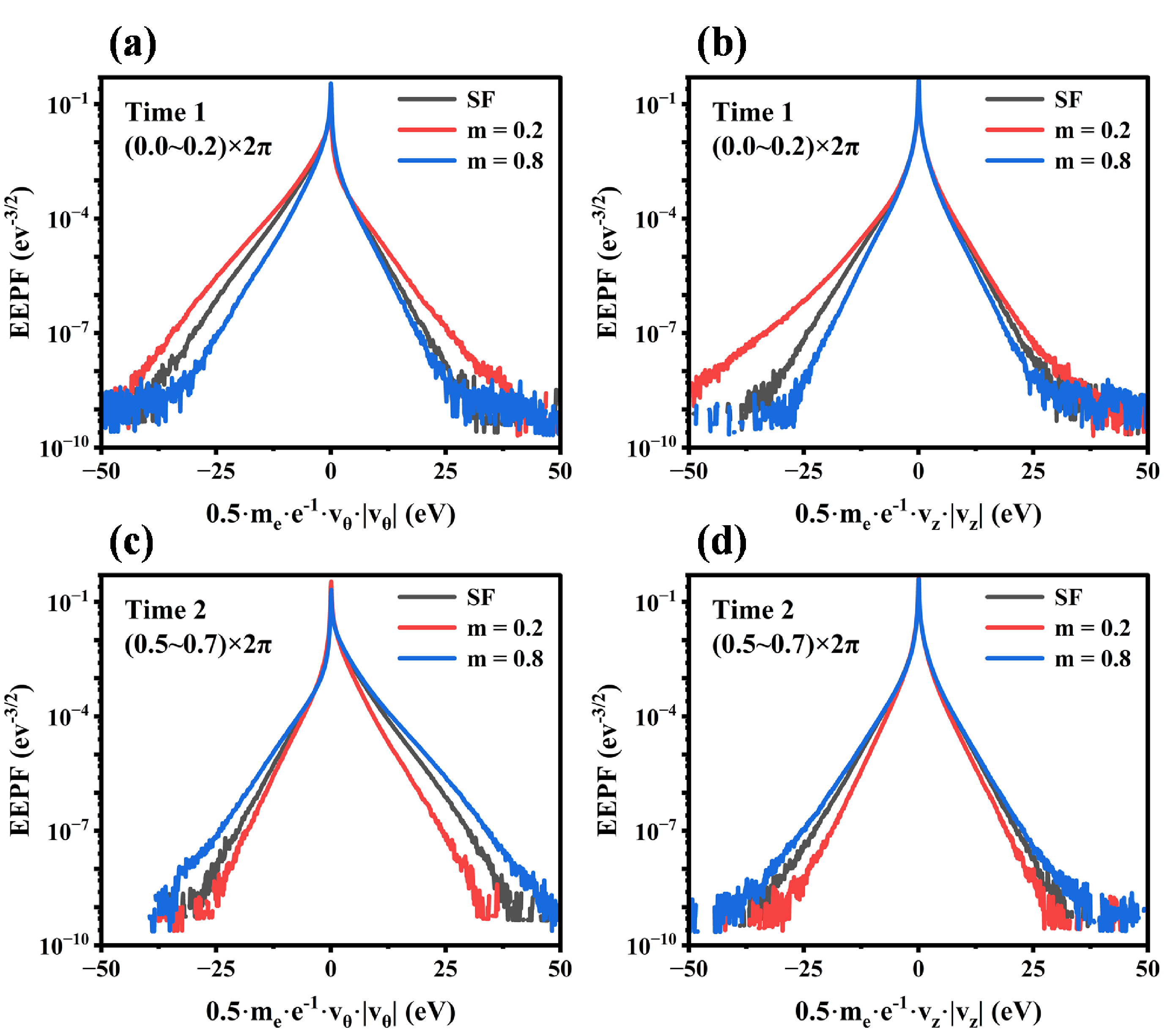}
	\end{center}
\caption{Direction-resolved EEPF in the sheath region 1, indicated in Fig. 1, and averaged over the phase intervals time 1 and 2, as indicated in Fig. 3. (a)(c) Azimuthal component ($v_\theta$), predominantly driven by the inductive field. (b)(d) Axial component ($v_z$), primarily governed by sheath dynamics.}
\label{eepf}
\end{figure}

\begin{figure}[htbp]
	\begin{center}
		\includegraphics[width=\linewidth]{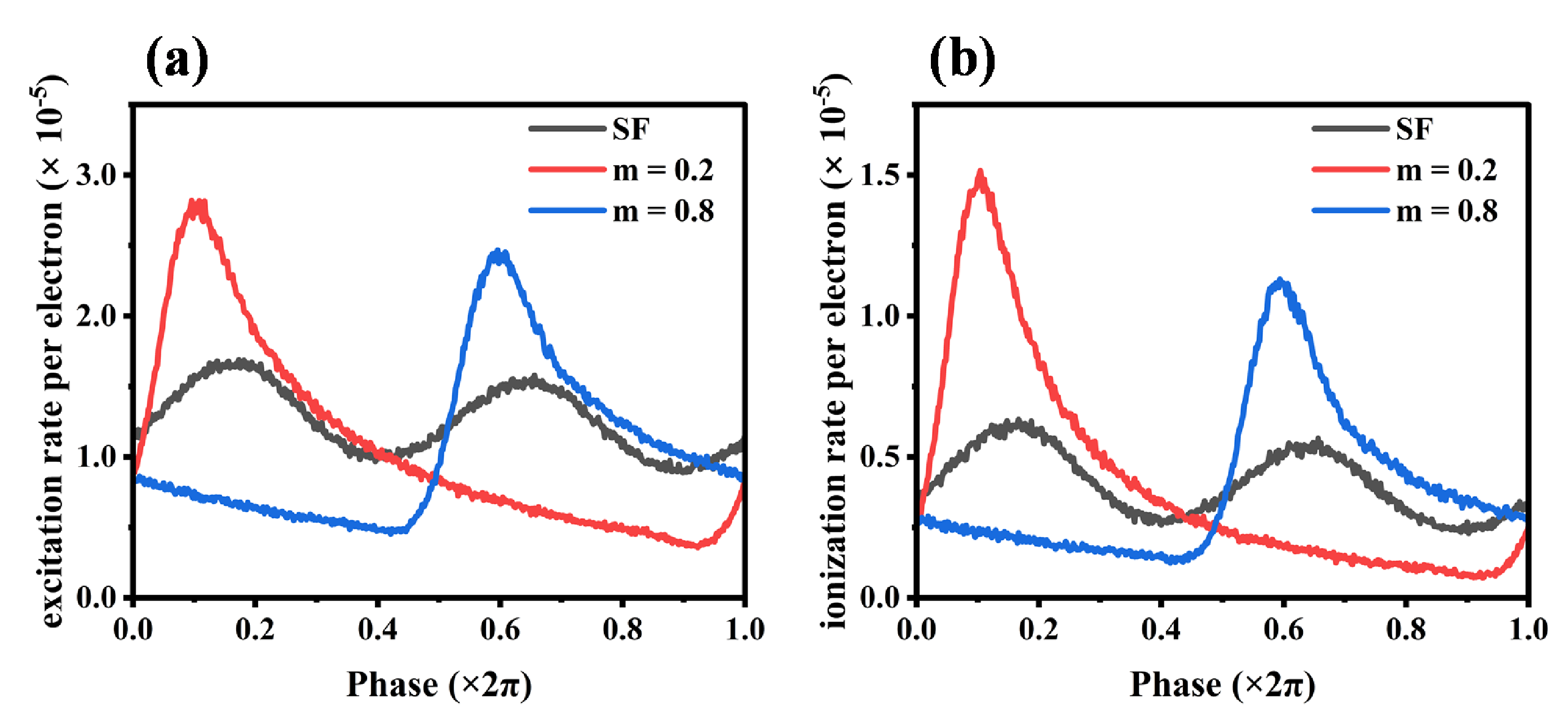}
	\end{center}
\caption{Temporal evolution of the spatially averaged (a) excitation rate per electron and (b) ionization rate per electron over one fundamental RF period.}
\label{ratio}
\end{figure}

In the single frequency base case, a standard 2D spatial distribution of the electron density is obtained with a maximum within the skin layer at a radial position of about 2/3 of the outer coil radius (Fig. \ref{neandphi}(a)). As a consequence of the capacitive coupling of the coil and the oscillating RF sheath on the plasma side of the dielectric window a DC voltage drop is established across the sheath to balance electron \cite{vahedi1997simultaneous} and ion fluxes \cite{lieberman2002analytical} locally at this window (Fig. \ref{neandphi}(b)). Since the dielectric surface is not equipotential, this DC bias exhibits pronounced radial non-uniformity, peaking beneath the high-voltage innermost coil turn. As shown in the left column of Fig. \ref{heating} at R = 3 cm, the electron power absorption happens primarily via the azimuthal induced electric field within the skin layer with weak contributions of the capacitive coupling of the coil, i.e., by the electrostatic axial electric field. The azimuthal electric field and, thus, the inductive electron power absorption are symmetric in time within the RF period leading strong power absorption and the presence of maxima of the density of high energy electrons ($\ge 16~\mathrm{eV}$) at two times within the RF period (time windows 1 and 2 in Fig. \ref{heating}), of which the first peak is slightly stronger due to the capacitive electron power absorption at that time. As shown in Figs. \ref{eepf} and \ref{ratio}, this leads to a specific EEPF shape and to temporally symmetric excitation and ionization rates per electron, that are inherently determined by the single frequency sinusoidal coil current waveform and cannot be controlled efficiently in standard single frequency ICPs.

As illustrated in Fig. \ref{model}(b), tailored sawtooth current waveforms, whose rise- and fall-times can be controlled by adjusting $m$ in equ. (\ref{eq_waveform}), are characterized by steep and low current gradients, $dI_{\mathrm{coil}}/dt$, at particular times within the fundamental RF period, i.e., the temporal symmetry inherent to single frequency ICPs is broken in this way. Given the relationship $E_{\theta} \propto -dI_{\mathrm{coil}}/dt$, this results in a strong inductive electric field within a narrow time window and a lower azimuthal field within a another longer time window per fundamental RF period. The self-consistent solution of the coil potential reveals that such rapid current transitions also induce large electromotive forces, leading to significant voltage fluctuations across the coil terminals, as shown in Fig. \ref{model}(c). This voltage drives intense capacitive heating, thereby establishing a hybrid heating regime where the inductive and capacitive modes are temporally synchronized.

Figures \ref{neandphi} (c) - (h) illustrate the consequences of using such sawtooth tailored current waveforms on the 2D spatially resolved and time averaged plasma density and potential. Using sawtooth current waveforms with steep gradients results in enhanced plasma densities. 

Columns 2 - 3 of Fig. \ref{heating} show that the presence of sawtooth coil current waveforms leads to a short phase of strong azimuthal electron power absorption per fundamental RF period due to a strong azimuthal electric field and to a longer separate phase of lower azimuthal electron power absorption due to a lower azimuthal electric field of opposite sign. The degree of this temporal asymmetry can be adjusted by tuning $m$, i.e., the steepness of the sawtooth coil current waveform. For a sawtooth-up waveform ($m = 0.2$) this time coincides with the time of maximum capacitive sheath expansion heating at the coil. This superposition of strong inductive and capacitive electron power absorption causes a strong single maximum of the density of energetic electrons at this time 1, as indicated in Fig. \ref{heating}. Using a sawtooth-down coil current waveform ($m = 0.8$) shifts the time of strong inductive electron power absorption to a later time 2 (see Fig. \ref{heating}) within the fundamental RF period, coinciding with the sheath collapse phase. This synchronization leads to a reduced sheath thickness, allowing the bulk plasma to expand closer to the upper dielectric window where the inductive field is stronger. This spatial proximity significantly enhances the heating efficiency, resulting in the minimum required coil current observed in Table \ref{power}. Thus, the maximum density of energetic electrons occurs at this later time and is weaker compared to the corresponding sawtooth-up coil current waveform.

This CWT control of the electric field and electron dynamics translates into effective control of the EEPF and plasma chemistry. Fig. \ref{eepf} shows the direction-resolved EEPF, spatially averaged over the region of maximum inductive electron power absorption (region 1 in Fig. \ref{model}) and time averaged over the first time window of strong inductive power absorption in the single frequency scenario (time 1 in Fig. \ref{heating}) in (a,b) and over the second time window of strong inductive power absorption in the single frequency scenario (time 2 in Fig. \ref{heating}) in (c,d), respectively. Results are shown separately for the azimuthal (a,c) and axial (b,d) directions. While the EEPFs remain temporally symmetric in the azimuthal direction for the single frequency case, this symmetry is broken and can be controlled by CWT. 

For sawtooth-up waveforms ($m = 0.2$) the high energy tail of the EEPF is enhanced in azimuthal and axial direction at negative velocities within the first time window, but not at positive velocities in the second time window (a,b). This asymmetry can be controlled by adjusting $m$ and is a consequence of the electric field and electron power absorption control by CWT. For sawtooth-down coil current waveforms ($m = 0.8$) the EEPF high energy tail is enhanced at positive azimuthal velocities within the second time window (c), but no such enhancement is found in axial direction (d), since the sheath is collapsing.

\begin{table}[htbp]
\centering
\renewcommand{\arraystretch}{2} 
\setlength{\tabcolsep}{5pt}      

\caption{Ratio of the spatially averaged total excitation rate to the total ionization rate, $\mathrm{Ratio} = \Sigma_{\text{exc}} / \Sigma_{\text{iz}}$, integrated over one fundamental RF period for different $m$.}
\begin{tabular}{lcccccc}
\hline \hline
\textbf{Case(m)} & $\mathbf{SF}$ & $\mathbf{0.2}$ & $\mathbf{0.4}$ &
$\mathbf{0.5}$ & $\mathbf{0.6}$ & $\mathbf{0.8}$        \\ \hline
Ratio & 3.115 & 2.623 & 3.067 & 3.085 & 3.066 & 2.718   \\ \hline 
\hline
\end{tabular}
\label{ratio01}
\end{table}

Such EEPF control by CWT provides control of the ratio of the total electron impact excitation and ionization rate, integrated over one RF period, as shown in Fig. \ref{ratio} and, thus, plasma chemistry, since these processes are sensitive to electrons at different energies. A lower ratio indicates that a larger fraction of electron energy is channeled into ionization rather than lower-threshold excitation pathways \cite{godyak2002electron}.
Regulating this ratio is of practical significance for process optimization: enhancing the ionization fraction allows for direct control over plasma density \cite{kushner2009hybrid}, while tuning the excitation level permits precise modulation of radical generation \cite{makabe2006plasma} and chemical reaction rates \cite{derzsi2013electron}.
As presented in Table \ref{ratio01}, the $m=0.2$ and 0.8 sawtooth current waveforms yield the lowest ratios of $2.623$ and $2.718$, compared to $3.115$ for the reference single frequency case. Consequently, higher electron densities are observed for the $m=0.2$ and $m=0.8$ cases (see Fig. \ref{neandphi}), primarily because more energy is utilized for ionization rather than excitation. For m = 0.8 this is achieved at lower capacitive power dissipation.

\section{Conclusions}

In summary, we have investigated the electron kinetics and plasma dynamics in an ICP reactor driven by tailored coil current waveforms using 2D PIC/MCC simulations. We revealed that sawtooth waveforms with steep and adjustable current gradients allow breaking the temporal symmetry of the electric field and electron power absorption dynamics inherent in classical single frequency ICPs. Through tailored and electrically controlled dynamics of the inductive azimuthal and the capacitive axial electric field dynamics, the high energy tail of the EEPF can be tailored and, thus, the ratio of high to lower energy electrons in the plasma can be controlled. This allows controlling the ratio of the total electron impact ionization and excitation rate indicating control over the plasma chemistry. In reactive gasses CWT is expected to provide selective control over radical and ion densities, each sensitive to the densities of electrons at different energies. This enhanced plasma control concept can be realized at any existing ICP reactor by upgrading the external power supply system based on existing multi-frequency impedance matchings for waveform tailoring \cite{Schmidt_2015,Schmidt_2018,Wang_2019} and without modifying the reactor itself.


\section*{Acknowledgments}
This work was supported by the National Magnetic Confinement Fusion Energy Research Project (2022YFE03190300), the National Natural Science Foundation of China (12275095, 11975174 and 12011530142) and the German Research Foundation (428942393). The authors gratefully acknowledge the valueable guidance of Dr. Shahid Rauf.

\section*{Data availability statement}
The data that support the findings of this study are available from the corresponding author upon reasonable request.

\nolinenumbers

\bibliography{ref}

\bibliographystyle{abbrv}

\end{document}